\documentclass[twocolumn,showpacs,preprintnumbers,amsmath,amssymb]{revtex4}

\usepackage{graphicx}
\usepackage{dcolumn}
\usepackage{bm}
\begin{document}

\title{Generation of atom-atom correlations inside and outside the mutual light cone}

\author{Juan Le\'on}
\email{leon@imaff.cfmac.csic.es}
\homepage{http://www.imaff.csic.es/pcc/QUINFOG/}

\author{Carlos Sab\'{i}n}%
 \email{csl@imaff.cfmac.csic.es}
\homepage{http://www.imaff.csic.es/pcc/QUINFOG/}
\affiliation{%
Instituto de F\'{i}sica Fundamental, CSIC
 \\
Serrano 113-B, 28006 Madrid, Spain.\\
}%


\date{\today}

\begin{abstract}
We analyze whether a pair of neutral two level atoms can become entangled in a finite time while they remain causally
disconnected. The interaction with the e. m. field is treated perturbatively  in the electric dipole approximation. We start
from an initial vacuum state and obtain the final atomic correlations for the cases where $n = 0, 1,$ or 2 photons are produced
in a time $t$, and also when the final field state is unknown. Our results show that correlations are sizeable inside and
outside the mutual light cone for $n= 1$ and 2, and also that quantum correlations become classical by tracing over the field
state. For $n = 0$ we obtain entanglement generation by photon propagation between the atoms, the correlations come from the
indistinguishability of the source for $n = 1$, and may give rise to entanglement swapping for $n = 2$.
\end{abstract}

\maketitle

\section{Introduction}

Quantum superposition and entanglement are the cornerstones lying at the foundations of quantum information and the principal
support of the new quantum technologies which are at different stages of conception and development at present. Putting
entanglement to work, enabling its use as a resource, is the key to the success of these technologies. Therefore, a complete
understanding of entanglement, necessary  at the fundamental level, is  also important for these developments to occur.
Entanglement can be envisaged in very different forms; it  originally appeared in quantum mechanics~\cite{epr} as a direct
connection between distant particles, a residue of past direct interaction between them \cite{schrodinger}. In quantum field
theory entanglement can be traced back to the non-locality of the vacuum state ~\cite{summerswerner,summerswernerII} or, simply,
to field propagation. Similar arguments operate for a lattice of coupled oscillators~\cite{plenioeisert}.

In this paper we analyze some features of entanglement generation closely related to the microscopic causality of quantum field
theory. Put in simple words, this work attempts to ascertain whether a pair of spatially separated parties (say, a pair of
neutral two level atoms $A$ and $B$) can get entangled in a finite time while they remain causally disconnected
\cite{fermi,hegerfeldt,powerthiru}. Each party interacts locally with the electromagnetic field, the carrier of the interaction.
We stress here that, as shown in \cite{powerthiru}, perturbation theory produces nonsignalling \cite{gisin} results for this
system and that the apparent causality violations come from the nonlocal specification of some final states. At first sight, the
question can be answered in the negative; if the parties remain causally separated from each other, they can not entangle.
However, the propagator $D(x,y)$ is finite even when $c (x-y)^0 < |\mathbf{x}-\mathbf{y}|$, and perhaps some correlations could
be exchanged between both parties~\cite{franson}. Alternatively, the correlations could be blamed on the preexisting
entanglement between different parts of the vacuum~\cite{reznik,reznikII}, which could be transferred to the atoms. Whatever the
point of view, correlations are exchanged through (time ordered) products, while only commutators are restricted to be causal
\cite{fransonII}. Our analysis can not sidestep that the role of the field goes beyond that of a mere carrier, quanta could be
absorbed from the field or escape in the form of photons \cite{cabrillo,lamata}. How does the entanglement between $A$ and $B$
depend on the state of the field? This question shapes our discussion below.

We will include in the final state all the perturbatively accessible field states, analyzing for each of them  the  correlations
in the reduced atomic state. We compute the entanglement measures for different values of $(x-y)^0$ and
$|\mathbf{x}-\mathbf{y}|$, that lie inside the atoms mutual light cone and beyond. The atomic state that results after tracing
over the states of the field is separable, which means, in the scheme of \cite{reznik}, that there is no transference of vacuum
entanglement, only classical correlations. In \cite{reznik}, these correlations become entanglement when a suitable time
dependent coupling with the scalar field is introduced. As pointed out in \cite{reznikII}, this would require an unrealistic
control of the atom-field interaction in the electromagnetic case that we are dealing with here. As an alternative way to
achieve entanglement between the atoms we consider a postselection process of the field states with $n=0, 1, 2$ photons. This is
a nonlocal operation and therefore entanglement generation is allowed. In \cite{franson}, only the vacuum case when
$|\mathbf{x}-\mathbf{y}|\gg c (x-y)^0 $ was analyzed, and no entanglement measures were considered. We get quantum correlations
for all the different field states. We also get useful hints on the nature of the correlations, whether they come from photon
exchange, source indistinguishability, etc.

\section{The model}

We will consider the field initially  in the vacuum state, including the cases with 0, 1 and 2 final photons to analyze
perturbatively the amplitudes and density matrices to order $\alpha$. We assume that the wavelengths relevant in the interaction
with the atoms, and the separation between them, are much longer than the atomic dimensions.  The dipole approximation,
appropriate to these conditions,  permits the splitting of the system Hamiltonian into two parts $H = H_0 + H_I$ that are
separately gauge invariant. The first part is the Hamiltonian in the absence of interactions other than the potentials that keep
$A$ and $B$ stable, $H_0 = H_A + H_B + H_{\mbox{field}}$. The second contains all  the interaction of the atoms with the field
\begin{equation}
H_I = - \frac{1}{\epsilon_0}\sum_{n=A,B} \mathbf{d}_n(\mathbf{x}_n,t)\,\mathbf{D}(\mathbf{x}_n,t) \label{a},
\end{equation}
where $\mathbf{D}$ is the electric displacement field, and $\mathbf{d}_n \,=\,\sum_i\, e\,\int d^3 \mathbf{x}_i\,
\langle\,E\,|\,(\mathbf{x}_i-\mathbf{x}_n)\,|\,G\,\rangle$ is the electric dipole moment of atom $n$, that we will take as real
and of equal magnitude for both atoms $(\mathbf{d}=\mathbf{d_A}=\mathbf{d_B})$, $|\,E\,\rangle$ and $|\,G\,\rangle$ being the
excited and ground states of the atoms, respectively.

In what follows we choose a system  given initially by the product state, $|\,\psi\,\rangle_0\,=\,
|\,E\,G\,\rangle\cdot|\,0\,\rangle$ in which atom $A$ is in the excited state $|\,E\,\rangle$, atom $B$ in the ground state
$|\,G\,\rangle$, and the field in the vacuum state $|\,0\,\rangle$. The system then evolves under the effect of the interaction
during a lapse of time $t$ into a state:
\begin{equation}
|\,\psi\,\rangle_t = T\,[e^{-i\, \int_0^t\,dt'\, H_I\,(t')/\hbar}\,]|\,\psi\,\rangle_0 ,\label{b}
\end{equation}
($T$ being the time ordering operator) that,  to  order $\alpha$, can be given in the interaction picture as
\begin{eqnarray}
|\mbox{atom}_1,\mbox{atom}_2,\mbox{field}\rangle_{t} =  ((1+a)\,|\,E\,G\rangle + b\,|\,G\,E\rangle)\,|\,0\rangle\nonumber\\
 +(u\,|\,G\,G\,\rangle+ v\,|\,E\,E\,\rangle)\,|\,1\,\rangle+
(f\,|\,E\,G\rangle+ g\,|\,G\,E\rangle)\,|\,2\rangle\  \label{c}
\end{eqnarray}
where
\begin{eqnarray}
a&=&\frac{1}{2}\langle0|T(\mathcal{S}_A^+ \mathcal{S}_A^- + \mathcal{S}_B^-\mathcal{S}_B^+)|0\rangle,\, b=
\langle0|T(\mathcal{S}^+_B
\mathcal{S}^-_A)|0\rangle\nonumber\\
u_A\,&=&\,\langle\,1\,|\, \mathcal{S}^-_A\,|\,0\,\rangle,\, v_B\,=\,\langle\,1\,|\,
\mathcal{S}^+_B\,|\,0\,\rangle \label{d}\\
f&=&\frac{1}{2}\langle2|T(\mathcal{S}_A^+ \mathcal{S}_A^- +\mathcal{S}_B^-\mathcal{S}_B^+)|0\rangle,\,
g=\langle2|T(\mathcal{S}^+_B \mathcal{S}^-_A)|0\rangle,\nonumber
\end{eqnarray}
being $\mathcal{S}\,=\,- \frac{i}{\hbar}  \int_0^t\, dt'\, H_{I}(t')=\mathcal{S}^{+}\, +\, \mathcal{S}^{-}$, $T$ being the time
ordering operator and $|\,n\,\rangle,\,\, n=\,0,\,1,\,2$ is a shorthand for the state of $n$ photons with definite momenta and
polarizations, i.e. $|\,1\,\rangle\,=\,|\mathbf{k},\, \mathbf{\epsilon}\,\rangle$, etc. The sign of the superscript is
associated to the energy difference between the initial and final atomic states of each emission. Among all the terms that
contribute to the final state (\ref{c}) only $b$ corresponds to interaction between both atoms, which is real interaction only
if $c\,t>L$ ($L$ being the interatomic distance). This would change at higher order in $\alpha$. Here, $a$ describes
intra-atomic radiative corrections, $u$ and $v$ single photon emission by one atom, and $g$ by both atoms, while $f$ corresponds
to two photon emission by a single atom. Details on the computations of these quantities would be given in Appendix A. In
Quantum Optics, virtual terms like $v$, $f$ and $g$, which do not conserve energy and appear only at very short times, are
usually neglected by the introduction of a rotating wave approximation. But here we are interested in the short time behavior,
and therefore all the terms must be included, as in \cite{powerthiru,milonni,compagnoI}. Actually, only when all these virtual
effects are considered, it can be said properly that the probability of excitation of atom $B$ is completely independent of atom
$A$ when $L>c\,t$ \cite{milonni, compagnoI} ($L$ being the distance between the atoms).

Finally, in the dipole approximation the actions $\hbar\, \mathcal{S}^{\pm}_N$  in (\ref{d}) reduce to
\begin{eqnarray}
\mathcal{S}^{\pm}\,=\,- \frac{i}{\hbar}  \int_0^t\, dt' \: e^{\pm i\Omega t'}\, \mathbf{d}\,\mathbf{E}(\mathbf{x},t')\label{e}
\end{eqnarray}
where  $\Omega = \omega_E -\omega_G$ is the transition frequency, and we are neglecting atomic recoil.  (\ref{e}) depends on the
atomic properties $\Omega$ and $\mathbf{d}$, and on the interaction time $t$. In our calculations we will take $(\Omega
|\mathbf{d}|/e c) = 5\,\cdot 10^{-3}$, which is of the same order as the 1s $\rightarrow$ 2p transition in the hydrogen atom,
consider $\Omega\,t \gtrsim 1$, and analyze the cases $(L/c\,t)\simeq 1$ near the mutual light cone, inside and outside.

Given a definite field state $|\,n\,\rangle$ the pair of atoms is in a pure two qubits state as shown in (\ref{c}). We will
denote these states by $|\,A,B,n\,\rangle$,  $\rho_{AB}^{(n)}\,=\,|A,B,n\rangle\,\langle A,B,n\,|$, and $\rho_{A}^{(n)}\,=\,Tr_B
\,\rho_{AB}^{(n)}$ in the following, and will compute the entropy of entanglement $\mathbb{S}^{(n)}$ \cite{bennett}:
\begin{equation}
\mathbb{S}^{(n)}=Tr\, \rho_{A}^{(n)}\log{\rho_{A}^{(n)}}\label{f}
\end{equation}
and the concurrence $\mathbb{C}^{(n)}$  \cite{wootters}:
\begin{equation}
\mathbb{C}^{(n)}= max\{0, \sqrt{\lambda_i}-\sum_{j\neq i}\, \sqrt{\lambda_j}\} \label{g}
\end{equation}
(being $\lambda_i$ the largest of the eigenvalues $\lambda_j$ ($j=1,...,4$) of
$[(\sigma_y\otimes\sigma_y)\rho*(\sigma_y\otimes\sigma_y)]\rho$) for them.

\section{The case with $n=0$}

We first consider the case $n=0$, where the field is in the vacuum state and, after (\ref{c}), the atoms are in the pure state
$((1\,+\,a)\,|\,E\,G\,\rangle + b\,|\,G\,E\,\rangle) / c_0$, where $c_0\,=\, \sqrt{|1\,+\,a|^2 \,+\, |b|^2}$ is the
normalization, giving a concurrence
 \begin{equation} \mathbb{C}^{(0)} \,=\, 2\,
|b|\,|\,1\,+\,a\,|/c_0^2\ .\label{h}
 \end{equation}
It is interesting to note that at lowest order the concurrence arises as an effect of the mutual interaction terms $b$ mediated
by photon exchange  or, in algebraic language, by the vacuum fluctuations. As expected,  at higher orders the radiative
corrections described by $a$ dress up these correlations. Analytic lowest order calculations (\cite{thiru}) showed that they can
persist beyond the mutual light cone, vanishing for $x = (L/c\,t) \rightarrow \infty$. We sketched in Fig. 1 the concurrence
$\mathbb{C}^{(0)}$ for $x$ around 1. Our computations were done for the illustrative case where both dipoles are parallel and
orthogonal to the line joining $A$ and $B$. We will adhere to this geometrical configuration for the rest of the letter. It
would correspond to an experimental set up in which the dipoles are induced by suitable external fields.  $\mathbb{C}^{(0)}$
shows a strong peak (of height 1) inside a tiny neighborhood of $x=1$. The features outside the mutual light cone are
$\vartheta(|\mathbf{d}|/e L)^2 \simeq 10^{-6}$ here, and could be larger if $\Omega t < 1$ entering into the Zeno region
(incidentally, $|b|\,\propto t^4$ for very small $t$ \cite{thiru}). Notice the change of behavior between the region where the
atoms are spacelike separated ($x<1$) and the region where one atom is inside the light cone of the other ($x>1$). This
qualitative treatment complements the quantitative one given in \cite{franson}.

\begin{figure}[h]
\includegraphics[width=0.5\textwidth]{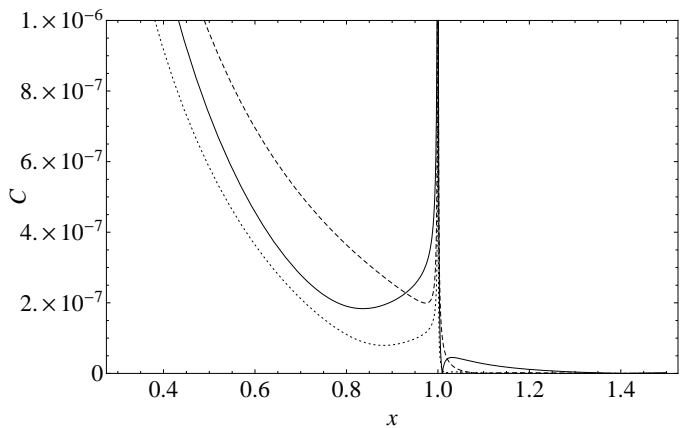}
\caption{Concurrence of the atomic state in the e.m. vacuum  $\rho_{AB}^{(0)}$ as a function of $x=(L/c\,t)$ for three values of
$z=(\Omega L/c)= 5$ (solid line), 10 (dashed line) and 15 (dotted line). The height of the peak is $\mathbb{C}^{(0)}=1$. }
\end{figure}
 The entropy of entanglement written in terms of the small quantity $\eta_0\,=\,(|b|/c_0)^2 \in (0,1)$ is
\begin{equation}
\mathbb{S}^{(0)}\,=\, - (1-\eta_0)\, \log (1-\eta_0)\,-\,\eta_0\,\log \eta_0\ ,\label{i}
\end{equation}
this is a positive quantity in $(0,1)$, which attains its maximum possible value $\mathbb{S}^{(0)}\,=\,1$ when the state is
maximally entangled at $\eta_0=\,0.5$. This is well within the small neighborhood of $x=1$ mentioned above. Radiative
corrections would shift the maximum to $|b|\,=\,|1 + a|$, so the entropy is sensitive to the Lamb shift when this contributes to
the dipole radiative corrections.

\section{Photon emission}
We now come to the case $n=1$, where the atoms excite one photon from the vacuum,  jumping to the state
$(\,u\,|\,G\,G\,\rangle\,+\,v\,|\,E\,E\,\rangle)/c_1$, (with $c_1 = \sqrt{|\,u\,|^2\,+\,|\,v\,|^2}$), during the time interval
$t$. The density matrix for this case contains the term $v\,u^* = Tr_1\,\langle\,1\,|\,
\mathcal{S}^+_B\,|\,0\,\rangle\,\langle\,1\,|\, \mathcal{S}^-_A\,|\,0\,\rangle^*\,=\,\langle\,0\,|\,
\mathcal{S}^+_A\,\mathcal{S}^+_B\,|\,0\,\rangle $, which we will call $l$ in the following, producing a concurrence
\begin{equation}
\mathbb{C}^{(1)} \,=\,2 |\,l\,|/c_1^2\ ,\label{j}
 \end{equation}
so, even if this case only describes independent local phenomena attached to the emission of one photon by either atom $A$ or
$B$, the concurrence comes from the tangling between the amplitudes $u$ and $v$ which have different loci. The state of the
photon emitted by $A$ and the state of $A$ are correlated in the same way as the state of the photon emitted by $B$ with the
state of $B$ are. These independent field-atom correlations are transferred to atom-atom correlations when we trace out a photon
line with different ends, $A$ and $B$, when computing $v\,u^*$. In fact, while $|u|^2$ and $|v|^2$ are independent of the
distance $L$ between the atoms,
\begin{equation}
 l= - {c d_A^id_B^j \over \hbar \epsilon_0}\, \{(\delta_{ij}-\hat{L}_i \hat{L}_j)M''(L)+(\delta_{ij}+\hat{L}_i \hat{L}_j){ M'(L)\over L} \}\label{k}
 \end{equation}
where

\begin{equation}
 M(L) =
 \int_{0}^{\infty}dk \,{\sin{k L}\over L}\,\delta^t(\Omega+c k)\,\delta^t(\Omega-c k)\label{l}
\end{equation}
which depends explicitly on $L$. Above we used $\delta^t(\omega)\,=\, \sin (\,\omega\,t/2)/(\pi \omega)$, which becomes
$\delta(\omega)$ in the limit $t\rightarrow \infty$. In Fig. 2 we represent $\mathbb{C}^{(1)}$ in front of $x={L/c\,t}$ for some
values of $z=\Omega L/c$. As the Figure shows, there may be  a significative amount of concurrence for all $x$, indicating that
$\rho^{(1)}$ is an entangled state inside and outside the mutual light cone. The peak at $x=1$ comes from the term with phase
$k(L - c t)$ that can be singled out from the linear combination of phasors in the integrand of (\ref{l}).
\begin{figure}[h]
\includegraphics[width=0.45\textwidth]{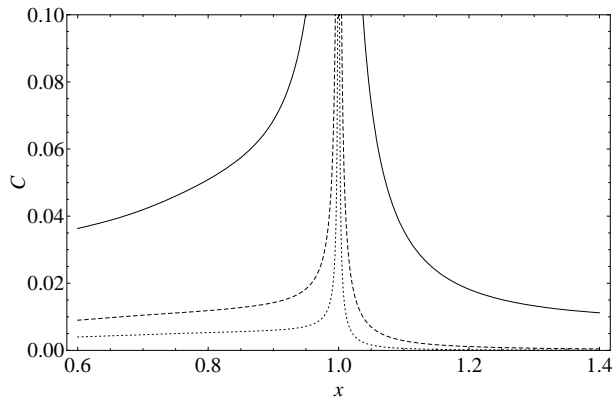}
\caption{Concurrence for one photon final state (\ref{j}) as a function of $x={L/c\,t}$ for three values of $z=\Omega L/c= 5$
(solid line), 10 (dashed line) and 15 (dotted line). Entanglement vanishes as $t\rightarrow\infty$ ($x\rightarrow0$ for a given
$L$) and is sizeable for $x>1$.}
\end{figure}

Here we have a lone photon whose source we can not tell. It might be $A$ or $B$, with the values of $l$ and $\mathbb{C}^{(1)}$
depending on their indistinguishability. Eventually, conservation of energy will forbid the process $G \rightarrow E + \gamma$
for large interaction times. Therefore, $v$, $l$ and $\mathbb{C}^{(1)}$ will vanish as $t$ grows to infinity ($x\rightarrow0$
for each value of $z$ in Fig. 2), as can be deduced from the vanishing of $\delta^t(\Omega+c k)$ for $t\rightarrow \infty$.

The entropy of entanglement gives an alternative description of the situation. Its computation requires tracing over one of the
parts $A$ or $B$, so no information is left in $\mathbb{S}^{(1)}$ about $L$, but it still gives information about the relative
contribution of both participating states $|E\,E\,\rangle$ and $|G\,G\,\rangle$  to the final state.  In terms of  $\eta_1\,=\,
|\,v\,|^2/{c_1}^2 \in (0,1)$, we have
 \begin{equation}
\mathbb{S}^{(1)}\,=\, - (1-\eta_1)\, \log (1-\eta_1)\,-\,\eta_1\,\log \eta_1 \label{m}
\end{equation}
Would not be for the difference between $\Omega + ck$ and  $\Omega -ck$, $v$ should be equal to $u$, $\eta_1\,= 0.5$, and
$\mathbb{S}^{(1)}$ would attain its maximum value. Not only this is not the case but, as said above, $v$ will vanish with time
and only $|G\,G\,\rangle$ will be in the final asymptotic state. Notice the result, indistinguishability was swept away because
for large $t$  we know which atom ($A$) emitted the photon. Therefore, the entropy will eventually vanish for large interaction
times.

There are two cases with $n=2$; one (with amplitude $f$) when both photons are emitted by the same atom, the other (with
amplitude $g$) when each atom emits a single photon. The final atomic state $(f\,|\,E\,G\,\rangle\,+\,g\,|\,G\,E\,\rangle)/c_2$,
with $c_2\,=\,\sqrt{|\,f\,|^2\,+\,|\,g\,|^2}$, is in the same subspace as for $n\,=\,0$. The normalization $c_2$ is
$\mathcal{O}(\alpha)$ like the expectation values $f$, $g$, so that all the coefficients in $\rho^{(2)}$ may be large. The
concurrence is $\mathbb{C}^{(2)}=2 |\,f\,g^*\,|/c_2^2$. Due to the tracing over photon quantum numbers,  $f\,g^*$ is a sum of
products containing not only  factors $u$ and $v$, but also $L$ dependent factors like $l$. The entropy $\mathbb{S}^{(2)}$ is
now given in terms of a parameter $\eta_2\,=\, |\,g\,|^2/c_2^2$. Notice that $|g|^2 = |u|^2 |v|^2+|l|^2$. Hence, both
$\mathbb{C}^{(2)}$ and $\mathbb{S}^{(2)}$, depend on $L$. This is different from the single photon case, where the only  $L$
dependence was in the coherences of $\rho^{(1)}_{AB}$, which did not feed into $\rho^{(1)}_{A}$. The correlations  came in that
case from the indistinguishability of the photon source. The case $n =2$ resembles that of the entanglement swapping paradigm
\cite{swapping}, where there are two independent pairs of down converted photons. Here we have two independent atom - photon
pairs. The swapping would arise in both cases from detecting one photon of each pair. But with the initial state we are
considering here, both $f$ and $g$ eventually vanish. More interesting would be the case with the initial atomic state
$|\,E\,E\,\rangle$, that we will analyze elsewhere.

\section{Tracing over the field}

We have seen that if the state of the field is defined, the atomic state is entangled inside and outside the light cone. But
what happens if the field state is ignored, that is, if we trace over the field degrees of freedom? Then the atomic state is
represented by the following density matrix (in the basis $\{|EE\rangle,|EG\rangle,|GE\rangle,|GG\rangle\}$):
\begin{eqnarray}
\rho_{AB}=\left( \begin{array}{c c c c}|v|^2&0&0&l\\0&|1+a|^2+|f|^2&(1+a)b^*+fg^*&0\\0& b(1+a)^*+f^*g&|b|^2+|g|^2&0\\
l^*&0&0&|u|^2
\end{array}\right)N^{-1}\label{n}
\end{eqnarray}
where $l=\langle0|\,\mathcal{S}_{A}^+\,\mathcal{S}_{B}^+\,|0\rangle$ was used again, and $N=|1+a|^2
+|b|^2+|u|^2+|v|^2+|f|^2+|g|^2$. It can be shown that the concurrence associated to this density always vanishes except for a
bounded range of small values of $x$.
\begin{figure}[h]
\includegraphics[width=0.45\textwidth]{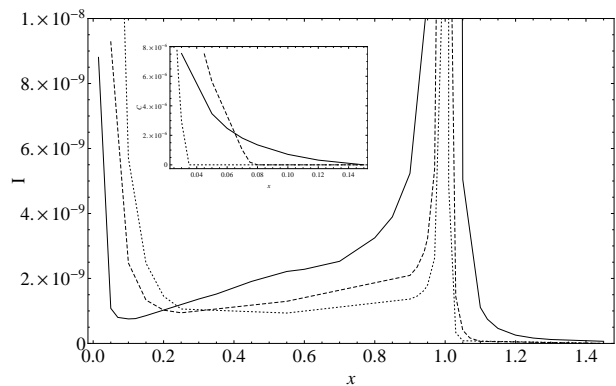}
\caption{Mutual information of $\rho_{AB}$ as a function of $x=L/c\,t$ for $z=\Omega L/c= 5$ (solid line), 10 (dashed line), 15
(dotted line). The inset shows the finite concurrences that are possible only for small values of $x$.}
\end{figure}
Beyond this range $\rho_{AB}$ is a separable state with no quantum correlations, either inside or outside the light cone. But
the atoms $A$ and $B$ are mutually dependent even for zero concurrence. Their mutual information $\mathbb{I}(\rho_{AB})=
\mathbb{S}(\rho_{A})+\mathbb{S}(\rho_{B})-\mathbb{S}(\rho_{AB})$, which measures the total correlations between both parties, is
completely classical in this case, but may be finite. We show this quantity in Fig. 4 for different values of $z$ with an inset
with the concurrence for small values of $x$.

\section{Conclusions}
In conclusion, we have studied the correlations between a pair of neutral two-level atoms that are allowed to interact with the
electromagnetic field, initially in the vacuum state. We have computed the concurrences that arise when the final state contains
$n=0,1,$ or 2 photons. They may be sizeable for $x$ small ($t\rightarrow\infty$ for a given $L$) and also around $x=1$. Only
in the case $n=0$ there are interactions between both atoms, generating an entanglement that persists asymptotically. We have
carefully taken into account all the terms contributing to the amplitude for finite time (they are $\propto t^4$, not $\propto
t^2$ as is sometimes assumed). A small amount of entanglement can be generated between spacelike separated parties due to the
finiteness of $b$ when $x>1$, but a change of behavior appears for $x<1$.  For $n=2$ the final atoms are in the same subspace
than for $n=0$. There are similar correlations that in this case can give rise to entanglement swapping, by measuring both
photons in a definite state for instance. Naturally, in this case entanglement may be sizeable for spacelike separated parties,
as here this is not related to any kind of propagation. Entanglement in the case with only a final photon ($n=1$)  comes from
the indistinguishability of the photon source. It will vanish asymptotically when, by energy conservation, only one atom ($A$ in
the present case) may emit the photon, and it is also sizeable when $x>1$. It is interesting how these correlations become
classical (except for small $x$) when the states of the field are traced over. We have shown through the mutual information the
residues of what were quantum correlations in the individual cases analyzed before.

\begin{acknowledgments}
We thank G. Garc\'ia-Alcaine, J. Garc\'ia-Ripoll, L. Lamata, S. Pascazio, D. Salgado and E. Solano for valuable discussions
concerning this work. This work was supported by Spanish MEC FIS2005-05304 and CSIC 2004 5 OE 271 projects. C.S. acknowledges
support from CSIC I3 program.
\end{acknowledgments}
\appendix*
\section{}
In this appendix, we will give details on the computations of the quantities of interest in the main text. We first start with
$a$ and $b$ in (\ref{d}). Both are a sum of second order transition amplitudes, which can be written as:
\begin{eqnarray}
(\frac {-i}{\hbar})^2\sum_{k}&\langle\,f|\,H_I\,|k\rangle\langle\,k|H_I|\,i\rangle\,\int^t_0 dt_1\int^{t_1}_0 dt_2 \nonumber\\
&e^{i(E_f-E_k)t_2/\hbar}e^{i(E_k-E_i)t_1/\hbar} \label{o}
\end{eqnarray}
being $E_f$, $E_k$ and $E_i$ the energies of the final $|\,f\rangle$, intermediate $|\,k\rangle$ and initial $|\,i\rangle$
states of the system, respectively. The sum over $k$ is a sum over all the possible intermediate states of the system which in
the case of fixed two-level atoms reduces to a sum over all the momenta and polarizations of the emitted photon. The time integrations in (\ref{o}) are
just
\begin{equation}
-\hbar^2(\frac{e^{i(E_f-E_i)t/\hbar}-1}{(E_f-E_k)(E_f-E_i)}-\frac{e^{i(E_k-E_i)t/\hbar}-1}{(E_f-E_k)(E_k-E_i)}) \label{p}
\end{equation}
The second term in (\ref{p}) is usually neglected, but give rise to a very different short time behavior \cite{thiru} ($\propto t^4$, not $\propto
t^2$). Therefore, it is of interest for our purposes.

In order to obtain $b$ we have to sum over the amplitudes for single photon emission at atom $A$ ($B$) followed by absorption at
atom $B$ ($A$). The case where a photon is emitted and absorbed by the same atom corresponds to $a$, that we will consider below. Using the mode expansion for the electric field:
\begin{eqnarray}
\mathbf{E}(\mathbf{x})&=&i\sqrt{\frac{\hbar\, c}{2\varepsilon_0\,(2\pi)^3}}\sum_{\lambda}\int d^3k
\sqrt{k}(e^{i\mathbf{k}\,\mathbf{x}}\mathbf{\epsilon}(\mathbf{k},\lambda)\,a_{k\lambda}\nonumber\\&-&
e^{-i\mathbf{k}\,\mathbf{x}}\mathbf{\epsilon}^*\,( \mathbf{k},\lambda)\,a^{\dag}_{k\lambda}), \label{q}
\end{eqnarray}
(with $[\,a_{k\lambda},\,a^{\dag}_{k'\lambda'}\,]=\delta^3(\mathbf{k}-\mathbf{k'})\,\delta_{\lambda\,\lambda'}$)  taking into
account (\ref{o}) and (\ref{p}), recalling that
$\sum_{\lambda}\mathbf{\epsilon}_i^*\,(\mathbf{k},\lambda)\,\mathbf{\epsilon}_j\,(\mathbf{k},\lambda)=\delta_{ij}-\hat{k_i}\hat{k_j}$,
and using the tabulated integrals that we list at the end of the appendix, a somewhat tedious although straightforward
computation leads to:
\begin{equation}
b= \frac{\alpha\,d^i d^j}{\pi\,e^2}(-\mathbf{\nabla}^2\delta_{ij}+\nabla_i\nabla_j)\,I \label{r}
\end{equation}
$\alpha$ being the fine structure constant and $I=I_+ + I_- $, which, in terms of $z=\Omega\,L/c$ and $x=L/c\,t$ are
\begin{eqnarray}
I_{\pm}={\frac{1\pm\frac{1}{x}}{2}}\{e^{i\,z}&[Ei(-iz)-Ei(-iz(1\pm1/x))]+\nonumber\\e^{-i\,z}&[Ei(iz)-Ei(iz(1\pm1/x))]\}
\label{s}
\end{eqnarray}
for $x>1$, $I$ having the extra term $i\pi(1-1/x)e^{-i\,z}$ for $x<1$. We use the conventions of \cite{bateman}. As noted in
\cite{thiru}, the non-zero contributions for $x>1$ come from the second term of (\ref{p}). We display here the results of the
derivatives in (\ref{r}) only for the particular case where the dipoles are parallel along the $z$ axis
($\mathbf{d}_A=\mathbf{d}_B=\mathbf{d}=d\,\mathbf{u}_z$) and the atoms are along the $x$ axis, corresponding to the physical
situation considered previously in, for instance, \cite{franson} and in this paper. Actually, $|\,E\,\rangle$ is a triply
degenerate state $|\,E\,,m\rangle$ with $m=0,\pm1$ and our scheme holds for a transition with $\Delta m=0$ \cite{milonniII}.
Another independent possibility would be to consider transitions with $\Delta m=\pm1$ that corresponds to
$\mathbf{d}=d\,(\mathbf{u}_x \pm i \mathbf{u}_y)/\sqrt{2}$ \cite{milonniII}. We find that:
\begin{eqnarray}
b&=-\frac{\alpha|\mathbf{d}|^2}{2\pi \,x\,L^2\,e^2}\{4 x (-1+\frac{(-2+x^2) \cos{\frac{z}{x}}}{-1+x^2})+e^{i z} [-2\,x\,z^2 Ei(-i z)\nonumber\\
&+\left(2+z \left(-2 i+(-1+x) z\right)\right)Ei(-\frac{i (-1+x) z}{x})\nonumber\\&+\left(-2+z (2 i+z+x z)\right) Ei(-\frac{i
(1+x) z}{x})]\nonumber\\&+2e^{-iz} \left(Ei(\frac{i (-1+x) z}{x})-Ei(\frac{i (1+x) z}{x})\right)+ \nonumber\\&z\,e^{-iz}
[-2\,x\, z Ei(i z)+\left(2 i+(-1+x) z\right) \nonumber\\&Ei(\frac{i (-1+x) z}{x})+(-2 i+z+x z) Ei(\frac{i (1+x)
z}{x})]\}\label{t}
\end{eqnarray}
for $x>1$, with the additional term $i\alpha\,e^{-iz}d^2\left(2+z\left(2i+(-1+x)z\right)\right)/(L^2\,x)$ for $x<1$. Please
notice that, for $x=1$, $Ei(0)=-\infty$, and therefore $b=\infty$, but then, recalling $(\ref{h})$, $\mathbb{C}^{(0)} \,=0$.

Now we come to $a$, which is the sum of the radiative corrections of atoms $A$ and $B$. As can be seen in the main text, $a$ appears in our results only as a higher order correction to $b$. Therefore, instead of finding an exact expression for it, we are mainly concerned with removing the divergencies. We followed the standard treatment (see,
for instance, \cite{cohentannoudji}) which is valid for the times $\Omega\,t>1$ we are considering. From (\ref{o}), it is
possible to arrive at:
\begin{equation}
\frac{-2\,i\,\alpha|\,d\,|^2\,t}{3\,\pi\,e^2\,c^2}\,\lim_{\epsilon\rightarrow0_+}\int_0^{\infty}d\omega\,\omega^3\,
({1\over{\Omega-\,\omega+i\,\epsilon}}-{1\over{\Omega+\,\omega-i\,\epsilon}}) \label{u}.
\end{equation}
Now, using in (\ref{u}) the identities
\begin{equation}
 {\omega^3\over\Omega\pm\, \omega}=\pm (\omega^2-\mp\,\Omega\,\omega+\Omega^2-{\Omega^3\over\Omega\pm\,\omega}), \label{v}
\end{equation}
the first term of (\ref{v}) cancels out the contribution of the Hamiltonian self-interaction terms \cite{cohentannoudji}, the
second is the state-independent contribution that can be absorbed in the definition of the zero of energy \cite{cohentannoudji},
the third cancels the counterterm coming from the mass renormalization \cite{cohentannoudji} and finally the last term has
logarithmic divergences and a cut-off, related with the fact that we are in the electric dipole representation could be imposed
at $t_{min}={a_0\over c}=1.76\cdot 10^{-19}\,s$. Please notice that the times relevant in our computations are of the order of
$t\cong {10\over\Omega}\approx 4\cdot 10^{-15}\,s$. Therefore:
\begin{equation}
a=\frac{2\,i\,\alpha\,|\mathbf{d}|^2\,z^3}{3\,\pi\,L^2\,e^2\,x}\,\ln{(|\frac{1-\frac{z_{max}}{z}}{1+\frac{z_{max}}{z}}|})],\label{w}
\end{equation}
with $z_{max}/z=c\,\Omega/a_0$.

Another quantity of interest is what we called $l$ in the main text and it is given by (\ref{k}) and (\ref{l}). Performing the
integration in (\ref{l}), we obtained $M(z,x)=M_+ (z,x)+ M_- (z,x)$, where:
\begin{eqnarray}
M_{\pm}(z,x) &=& \frac{e^{i\frac{z}{x}}}{4\pi^2\,z}\{\sin{(z(1\pm\frac{1}{x}))}[ci(z)-ci(z(1\pm\frac{1}{x}))]-\nonumber\\
&{}&\cos{(z(1\pm\frac{1}{x}))}[si(z)-si(z(1\pm\frac{1}{x}))]\}\label{x}
\end{eqnarray}
The derivatives in (\ref{k}) were performed in the same particular situation as in $b$.

$|u|^2=\langle0|\mathcal{S}^+_A \mathcal{S}^-_A|0\rangle,\,$ and $|v|^2=\langle0|\mathcal{S}^-_B\mathcal{S}^+_B|0\rangle,\,$ are
just the two terms that contribute to $a$ without the time ordering and therefore their divergencies are removed by
the application of (\ref{u}). Taking this into account, we obtain:
\begin{eqnarray}
&|u|^2&=\frac{2\alpha|d|^2 z^2}{3\pi\,e^2\,L^2} (-2+\pi \frac{z}{x}+2 \cos(\frac{z}{x})+2
(\frac{z}{x})si(\frac{z}{x}))\nonumber\\
&|v|^2&=\frac{2\alpha|d|^2 z^2}{3\pi\,e^2\,L^2} (2+\pi \frac{z}{x}-2 \cos(\frac{z}{x})-2 (\frac{z}{x})si(\frac{z}{x}))\ \ \ \ \
\ \ \ \ \label{y}
\end{eqnarray}
$f$ and $g$ can be written in terms of previously computed quantities, taking into account that:
\begin{eqnarray}
f&=&\theta(t_1-t_2)\,(\,u_A\,(t_1)v'_A(t_2)\,+\,v'_A\,(t_1)u_A\,(t_2)\nonumber\\
&+&\,u_B\,(t_1)v'_B\,(t_2)+\,u'_B\,(t_1)v_B(t_2)\,)\nonumber\\
g&=&u_A\,v'_B\,+\,u'_A\,v_B, \label{z}
\end{eqnarray}
being $v_A\,=\,\langle\,1\,|\, \mathcal{S}^+_A\,|\,0\,\rangle,\,$ and $u_B\,=\,\langle\,1\,|\, \mathcal{S}^-_B\,|\,0\,\rangle$.
The primes are introduced to label two different single photons. Therefore, in the computation of $|\,f\,|^2$,
$|\,g\,|^2$ and $f\,g^*$ we will only need, besides $|\,u\,|^2$, $|\,v\,|^2$ and $l$, the following:
\begin{eqnarray}
v_A\,u_A^* &=&v_B\,u_B^*=\frac{2\,\alpha|\,d\,|^2\,z^2}{3\,\pi\,L^2}\,e^{i\,\frac{z}{x}}\,\sin{\frac{z}{x}} \nonumber\\
v_A\,v_B^*&=& \frac{\alpha\,d^i d^j}{\pi}(-\mathbf{\nabla}^2\delta_{ij}+\nabla_i\nabla_j)\,I' \label{zz}
\end{eqnarray}
being $I'=I'_+ + I'_- $, with:
\begin{eqnarray}
I'_{\pm}&=&{\frac{1\pm\frac{1}{x}}{2}}\{e^{i\,z}\,Ei(-iz(1\pm\frac{1}{x}))+e^{-i\,z}\,Ei(iz(1\pm\frac{1}{x}))\}\nonumber\\
&-&e^{\mp\,i\,z}\,Ei(\pm\,i\,z)\} \label{zzz}
\end{eqnarray}
and $u_A\,u_B^*=v_A\,v_B^*$ when $x>1$, with the additional term $-2\,\pi\sin{z}\,(1-1/x)$ when $x<1$. Again the derivatives
were performed as in $b$ and $l$.

The following integrals are useful to obtain the results of this appendix \cite{bateman}:
\begin{eqnarray}
\int_0^\infty\,d\omega\,\frac{e^{\pm\,i\,\omega\,\gamma}}{\omega\,+\,\beta}&=&-e^{\mp\,i\,\gamma\,\beta}\,Ei(\pm\,i\,\gamma\,\beta)\label{zzzz}\\
\int_0^\infty\,d\omega\,\frac{e^{\pm\,i\,\,\omega\,\gamma}}{\omega\,-\,\beta}&=&-e^{\pm\,i\,\gamma\,\beta}\,(Ei(\mp\,i\,\gamma\,\beta)\,\mp\,i\,\pi),
\nonumber
\end{eqnarray}
with $a>0$, $arg\, \beta\leq\pi$.

\end{document}